| Title | **Cyto- and bio-compatibility assessment of plasma-treated polyvinylidene fluoride scaffolds for cardiac tissue engineering** |
|---|---|
| Authors | Maria Kitsara[1]*, Gaëlle Revet[1], Jean-Sébastien Vartanian-Grimaldi[1], Alexandre Simon[1], Mathilde Minguy[1], Antoine Miche[2], Vincent Humblot[2,3], Thierry Dufour[4]*† and Onnik Agbulut[1]† |
| Affiliations | [1]UMR CNRS 8256, INSERM ERL 1164, Biological Adaptation and Ageing, Institut de Biologie Paris-Seine, Sorbonne Université, Paris, France |
| | [2]UMR CNRS 7197, Laboratoire de Réactivité de Surface, Sorbonne Université, Paris, France |
| | [3]UMR 6174 CNRS, FEMTO-ST Institute, Université Bourgogne Franche-Comté, Besançon, France |
| | [4]UMR CNRS 7648, Laboratoire de Physique des Plasmas, Sorbonne Université, Paris, France |
| | † Co-senior authorship |
| Correspondence | Maria Kitsara, kitsara.m@gmail.com |
| | Thierry Dufour, thierry.dufour@sorbonne-universite.fr |
| | Onnik Agbulut, onnik.agbulut@sorbonne-universite.fr |
| Ref. | Front. Bioeng. Biotechnol., Vol. 4, pp. 18 (2022) |
| DOI | https://doi.org/10.3389/fbioe.2022.1008436 |
| Summary | As part of applications dealing with cardiovascular tissue engineering, drop-cast polyvinylidene fluoride (PVDF) scaffolds have been treated by cold plasma to enhance their adherence to cardiac cells. The scaffolds were treated in a dielectric barrier device where cold plasma was generated in a gaseous environment combining a carrier gas (helium or argon) with/without a reactive gas (molecular nitrogen). We show that an Ar-N$_2$ plasma treatment of 10 min results in significant hydrophilization of the scaffolds, with contact angles as low as 52.4° instead of 132.2° for native PVDF scaffolds. Correlation between optical emission spectroscopy and X-ray photoelectron spectroscopy shows that OH radicals from the plasma phase can functionalize the surface scaffolds, resulting in improved wettability. For all plasma-treated PVDF scaffolds, the adhesion and maturation of primary cardiomyocytes is increased, showing a well-organized sarcomeric structure (α-actinin immunostaining). The efficacy of plasma treatment was also supported by real-time PCR analysis to demonstrate an increased expression of the genes related to adhesion and cardiomyocyte function. Finally, the biocompatibility of the PVDF was studied in a cardiac environment, after implantation of acellular scaffolds on the surface of the heart of healthy mice. Seven and 28 days after implantation, no exuberant fibrosis and no multinucleated giant cells were visible in the grafted area, hence demonstrating the absence of foreign body reaction and the biocompatibility of these scaffolds. |
| Keywords | Cardiac tissue engineering, scaffolds, plasma treatment, surface functionalization, biocompatibility, polyvinylidene fluoride |

# I. Introduction

The objective of cardiovascular tissue engineering is to develop therapeutic options for structural, functional and vascular heart diseases, especially through the innovation of contractile tissues that can replace missing or damaged myocardial tissues. Among the various cardiovascular tissue engineering strategies investigated so far, biomaterials and scaffolds represent a major component either alone or combined with cells and/or bioactive molecules (Camman et al., 2022; Kitsara et al., 2022). The materials of these scaffolds must provide optimal support but also ensure cellular functions (reseeding, proliferation, differentiation, functional integration).

To date, two main approaches have been considered: the development of natural or synthetic materials and the development of native tissue derived acellular scaffolds, e.g., decellularized extracellular matrix derived from myocardium (Wang et al., 2012). Although the use of acellular scaffolds for transplantation should theoretically not threaten the recipient's immune system, experimental data have revealed the existence of significant immune responses to allogeneic and xenogeneic transplanted scaffolds (Bilodeau et al., 2020). Besides, the current processes of decellularization and recellularization of decellularized extracellular matrix still face many challenges (e.g., balance between cell removal and preservation of the extracellular matrix, efficient recellularization of decellularized extracellular matrix to achieve homogeneous cell distribution) (Kc et al., 2019). For these reasons, synthetic materials remain a serious alternative for cardiovascular tissue engineering.

Although the development of synthetic materials is an approach facing its own difficulties (potential inflammatory and immune reactions, mechanical flexibility, control of degradation rate), it relies on the use of polymeric materials that are easier to process than decellularized extracellular matrix, including polyglycolic acid (PGA), polylactic acid (PLA), polyester (urethane urea) (PEUU), polycaprolactone (PCL), polyethylene oxide (PEO), polyvinyl alcohol (PVA), and polyglycerol sebacate (PGS) (Wang et al., 2012). Among these polymer materials, PLA shows interesting features, especially its piezoelectric activity. Since the myocardium is an electroactive tissue that can contract rhythmically, the development of piezoelectric scaffolds is particularly attractive (Hitscherich et al., 2016). Indeed, minute deformations (e.g., pressure, strain) of these materials can generate electrical charges, hence converting mechanical energy into electrical energy without requiring any power sources or electrodes. For this reason, PLA-based polymers can be investigated for cardiovascular tissue engineering applications, e.g., core/shell fibers consisting of PGS as core material and PLA as shell material fabricated via coaxial electrospinning for the reconstruction of broken myocardium (Ravichandran et al., 2012). Random or aligned nanofiber scaffolds based on PLA/chitosan have also been designed using electrospinning methods to improve the formation of extracellular matrix and cell–scaffold interaction (Liu et al., 2017). However, PLA piezoelectric activity as well as its mechanical strength were much lower than that of polyvinylidene fluoride (PVDF) which therefore appears as a good–surprisingly poorly investigated–candidate for cardiovascular tissue engineering applications (Fryczkowski et al., 2013; Hitscherich et al., 2016; Adadi et al., 2020). Here, we propose to use PVDF scaffolds fabricated by drop casting, because of the ease and low cost of this method which does not require special equipment.

PVDF is a thermoplastic semi-crystalline polymer which, in more of being cheap and easy-to-process, exhibits attractive electroactive properties. Depending on the conformations of its macromolecular chain, it can crystallize into five different types of phases, where α, β and γ are the most important: α is non-polar, while β and γ are polar (Fakhri et al., 2016). Since α remains the most common form of PVDF, several processes have already been explored to transform this non-polar phase into polar. These processes include mechanical stretching, polarization and electrospinning (Kitsara et al., 2015; Kitsara et al., 2017; Deng and Chen, 2021; Angel et al., 2022; Li et al., 2022; Song et al., 2022; Yuan et al., 2022).

In our previous study, we showed that drop-cast PVDF scaffolds consist of globular particles interconnected with thin nanofibrils that form a high-porosity structure (Kitsara et al., 2019). This globular structure was obtained for samples dissolved in dimethylformamide (DMF) (Gregorio Jr and Ueno, 1999), while the formation of pores was obtained in the co-presence of acetone which is a low-boiling solvent with higher evaporation rate. Regarding piezoelectricity, we proved by X-ray diffraction (XRD) and Fourier transform infrared spectroscopy (FTIR) analyses that the scaffolds fabricated by drop-casting exhibited piezoelectric properties, mainly attributed to the γ-phase. Indeed, considering the drying of the drop-cast scaffolds at room temperature in a fume hood (to accelerate the solvent evaporation) and considering the higher proportion of volatile acetone compared to DMF in the final solution, it was found that the transition from the α-phase to the γ-phase was the most important (Kitsara et al., 2019). Now that the polarization of PVDF under its γ-phase is under control, it is essential to understand how the wetting properties of PVDF, a hydrophobic polymer, can be improved to achieve satisfactory cell adhesion to its surface (Dufour et al., 2013; Lech et al., 2020).

In this study, we propose to improve the adhesion of cardiac cells on drop-cast PVDF scaffolds using an atmospheric pressure plasma process. To this end, PVDF scaffolds were exposed to different plasma gas mixtures to modify their surface chemical properties. Then, the different plasma-treated scaffolds were cellularized with cardiomyocytes from neonatal rat hearts. Structural analysis of these cardiomyocytes allowed to determine the optimal plasma conditions to treat the scaffolds. In addition, biocompatibility was assessed in a cardiac environment by implanting acellular PVDF scaffolds into the hearts of healthy mice.

# II. Material & Methods

## II.1. Drop-casting of the PVDF scaffolds

PVDF powder (M.W. 534 K), DMF and acetone were purchased from Sigma-Aldrich. PVDF solutions of 3% w/v (weight per volume) and 5% w/v in DMF/acetone (40/60) were prepared by simultaneously agitating and heating at 60°C for at least 2 h using magnetic stirrer hotplate (Thermo Fisher Scientific, Saint-Herblain, France). Then a drop of each solution was deposited on glass lamella of 13 mm diameter. The drop-cast scaffolds were let to dry under a fume hood for 2 days at room temperature.

## II.2 Scaffolds treatment with cold plasma

The scaffolds were exposed to micro-discharges of cold plasma in a dielectric barrier device (DBD) operating at atmospheric pressure. This device comprised a quartz tube showing a rectangular section of 30 mm by 10 mm with a length of 30 cm. It was completed by two electrodes 10 cm long: an electrode polarized to high voltage (sine, 8 kV, 600 Hz) and a counter-electrode connected to a capacitor (100 pF) that was connected to the ground, as sketched in Figure 1. The area of the interelectrode gap was large enough to simultaneously treat 6 scaffolds. The DBD was supplied with a carrier gas (helium or argon) at a flow rate of 2 slm with/without reactive gas (molecular nitrogen) on a range between 0 and 20 sccm. All plasma treatments were operated during 10 min.

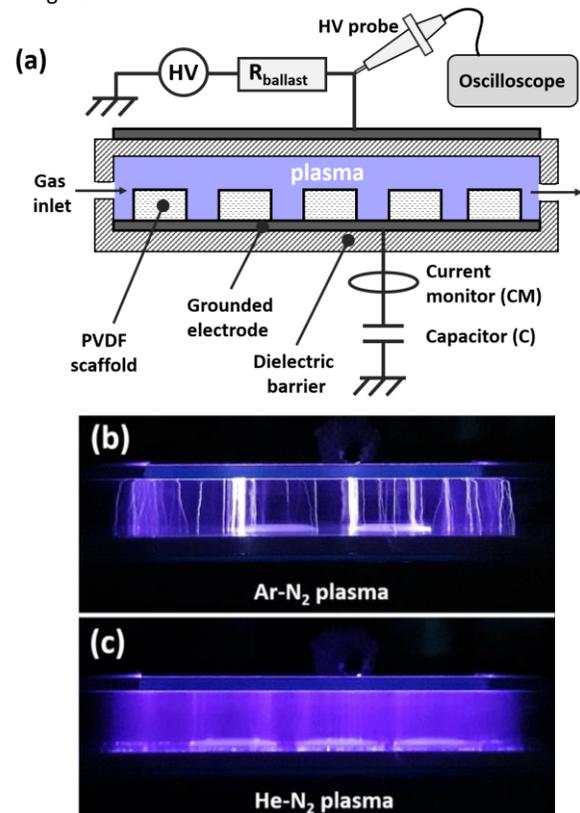

**Figure 1. Dielectric barrier device (DBD) set-up. (a) Schematic representation of the DBD. Images during operation of (b) Ar-N₂ plasma and (c) He-N₂ plasma.**

## II.3 Plasma diagnostics

The plasma phase was analyzed using an optical emission spectrometer (SR-750-B1-R model from Andor company) equipped with an ICCD camera (Istar DH340T model from Andor company). The spectrometer operated in the Czerny Turner configuration with a focal length of 750 mm while diffraction was achieved with a 1200 grooves. $mm^{-1}$ grating in the visible range. The following parameters were selected for all experiments: intensification factor = 4000, exposure time = 50 ms, number of accumulations = 200. Electrical measurements were carried out using an analog oscilloscope (Wavesurfer 3054, Teledyne LeCroy), a high voltage probe (Tektronix P6015A 1000:1) and a current monitor (Pearson, 2877), as sketched in Figure 1. Plasma electrical power was measured by plotting the Lissajous curve of the DBD (Dufour and Gutierrez, 2021). A current transformer (Pearson company, model 2877) is also utilized to characterize the distribution of the micro-discharges.

## II.4 Scaffolds' surface characterization

Scanning electron microscope (SEM) was used to visualize the structure of the scaffolds using the system Gemini SEM 500 from Zeiss. The samples were sputtered with a thin platinum layer prior to SEM observation using a high vacuum sputter coater (ACE600, Leica). Images at different magnifications were obtained on at least five different areas in the same sample and in three different samples with the same conditions. The images were performed by applying a beam voltage of 3 kV and using InLens detector.

Water contact angle (WCA) measurements were performed using the Drop Shape Analyzer 30 system from Krüss. The measurements were achieved at room temperature (20°C) following the Sessile drop method. Contact angle analyses were performed every 10 s after placing a drop of deionized water of volume 1 µL on the scaffold. WCA values were obtained by averaging at least three measurements in different areas of the sample surface.

X-ray photoelectron spectroscopy (XPS) analyses were performed using an Omicron Argus X-ray photoelectron spectrometer, equipped with a monochromated Al $K_\alpha$ radiation source (hν = 1486.6 eV) and a 300W electron beam power. The emission of photoelectrons from the sample was analyzed at a takeoff angle of 90° under ultra-high vacuum conditions (≤10−10 Torr). Spectra were carried out with a 100 eV pass energy for the survey scan and 20 eV pass energy for the C 1s, O 1s and F 1s regions. Binding energies were calibrated against the C1s binding energy at 284.8 eV and element peak intensities were corrected by Scofield factors (Scofield, 1976). The peak areas were determined after subtraction of a linear background. The spectra were fitted using Casa XPS v.2.3.15 software (Casa Software Ltd., United Kingdom) and applying a Gaussian/Lorentzian ratio G/L equal to 70/30.

## II.5 Isolation and cell culture of neonatal rat cardiomyocytes

Newborn rat primary cardiomyocytes were obtained from 1-day-old Wistar RjHan rat (Janvier Labs, Saint-Berthevin, France). The hearts were cut in small pieces and the cardiac cells were isolated, using a neonatal heart dissociation kit (Miltenyi Biotec, Paris,

France) according to the manufacturer's instructions. Cardiomyocytes were purified by depletion of non-target cells (such as fibroblasts, endothelial cells, blood cells) using a neonatal cardiomyocyte isolation kit (Miltenyi Biotec, Paris, France). $5 \times 10^5$ cardiomyocytes were seeded on each PVDF scaffold mounted in CellCrown™48 inserts (Scaffdex, Tampere, Finland), each of these inserts having an inner surface area of 0.38 $cm^2$. Since drop-cast PVDF scaffolds exhibited globular morphology and high porosity, their surface areas are not planar but rather tridimensional structures, which precludes any accurate estimation of a real cell density. The cardiomyocytes were cultured with DMEM medium (with 4.5 g/L D-Glucose and without Na Pyruvate) supplemented with L-glutamine, penicillin-streptomycin, 10% horse serum and 5% foetal bovine serum at 37°C and 5% $CO_2$ for 3 and 6 days. All animal studies were approved by our institutional Ethics Committee "Charles Darwin" and conducted according to the French and European laws directives, and regulations on animal care (European Commission Directive 2010/63/EU) under the license B75.13.20.

## II.6 Immunostaining and morphology analysis of primary cardiomyocytes

The primary cardiomyocytes were washed with PBS and fixed with 4% paraformaldehyde solution. After 3 washing steps, they were incubated with 5% bovine serum albumin for 1 h. Then, the cells were incubated for 90 min at room temperature with antibodies against α-actinin (mouse IgG1, clone EA-53, dilution 1:300, Sigma-Aldrich, #A7811, Saint-Quentin-Fallavier, France). After 3 washes in PBS, primary cardiomyocytes were incubated in the presence of the secondary antibody (Alexa 488 goat-anti-mouse, dilution 1:1000; Life Technologies). The scaffolds were mounted with mowiol containing 5 µg/ml Hoescht 33,342 (Life Technologies). Images were captured using a motorized confocal laser scanning microscope (Leica TCS SP5). Cardiomyocyte surface area and perimeter, cardiomyocyte length-to-width ratio and roundness index were measured using ImageJ software as described in (Flaig et al., 2020).

## II.7 Relative quantification of gene expression by real-time PCR

Total RNA was extracted from the primary cardiomyocytes using TRIzol® (Thermo Fisher Scientific, Saint-Herblain, France) following the manufacturer's instructions. From 500 ng of extracted RNA, the firststrand cDNA was then synthesized using a RevertAid First Strand cDNA Synthesis Kit (Thermo Fisher Scientific, Saint-Herblain, France) with random hexamers according to the manufacturer's instructions. Using the Light Cycler® 480 system (Roche Diagnostics), the reaction was carried out in duplicate for each sample in a 6 µL reaction volume containing 3 µl of SYBR Green Master Mix, 500 nM of the forward and reverse primers each and 3 µl of diluted (1:25) cDNA. The thermal profile for the SYBR Green qPCR was 95°C for 8 min, followed by 40 cycles at 95°C for 15 s, 60°C for 15 s and 72°C for 30 s. To exclude PCR products amplified from genomic DNA, primers were designed, when possible, to span one exon-exon junction. The mean gene expression stability of 3 genes, actin beta (Actb), beta-2-microgobulin (B2m) and NUBP iron-sulfur cluster assembly factor 1 (Nubp1), were used as the reference transcripts. Data were collected and analyzed using the LightCycler® 480 software release 1.5.0 (Roche Diagnostics). Primers sequences used in this study are presented in Supplementary Table S1.

## II.8 Assessment of cardiac and adhesion index

The cardiac index was constructed using the fold change relative to control (non-treated PVDF) in quantitative PCRbased expression of α-actinin (Actn2), myosin heavy chain 7 (Myh7), phospholamban (Pln), ATPase Sarcoplasmic/Endoplasmic Reticulum $Ca^{2+}$ Transporting 2 (Atp2a2) and alphaB-crystallin (Cryab). The adhesion index was constructed using the fold change relative to control (non-treated PVDF) in quantitative PCR-based expression of cadherin-2 (Cdh2), cadherin-13 (Cdh13), collagen-type 1-alpha 1 (Col1α1), extracellular matrix protein1 (Ecm1), elastin (Eln), laminin subunit alpha 4 (Lama4), laminin subunit beta 2 (Lamb2) and laminin subunit gamma 1 (Lamc1). Each index value was obtained by performing a two-step calculation: 1) normalization of the PCR results as the ratio of the mean of "plasma-treated PVDF scaffolds" to the mean of "non-treated PVDF scaffolds'', 2) averaging the ratios for each of the four plasma conditions. This calculation explains why the control has an index value always equal to 1.

For more information, Supplementary Tables S2, S3 of Supplementary Materials reports all genes considered separately (Supplementary Table S2 indicates the values after normalization and Table S3 indicates the values before normalization).

## II.9 Epicardial grafting in mice

The biocompatibility assessment was performed on n = 37 mice (8-week-old female Swiss mice from Janvier Labs, Saint-Berthevin, France). All procedures were performed in accordance with national and European legislations, in accordance with the Public Health Service Policy on Human Care and Use of Laboratory Animals under the license B75.13.20. All animal studies were approved by our institutional Ethics Committee "Charles Darwin" (Permit number: 4370) and conducted according to the French and European laws, directives, and regulations on animal care (European Commission Directive 86/609/EEC). All animals underwent left lateral thoracotomy after intraperitoneal ketamine (100 mg/kg; Merial, France)-xylazin (10 mg/kg; Bayer, France) anesthesia and tracheal ventilation. Analgesia was performed for 2 days after surgery with a 2 mg/kg intraperitoneal injection of profenid® (Merial). Acellular PVDF scaffolds were sutured (surface approximately 60 mm$^2$) on the surface of the ventricle of the mice. To assess the biocompatibility of the scaffolds, two series of experiment were performed. First, non-treated PVDF scaffolds were sutured on the surface of the ventricle of the mice and the animals were sacrificed by cervical dislocation 7 days (n = 6) and 28 days (n = 6) after implantation. Second, Ar-N$_2$ (n = 5), He-N$_2$ (n = 6) plasma treated PVDF scaffolds were sutured on the surface of the ventricle of the mice and the animals were sacrificed by cervical dislocation 7 days after implantation. Besides, n = 5 mice served as sham-operated controls. After sacrifice, hearts were removed and processed for histological analysis.

## II.10 Tissue processing and morphological analysis

For histological analysis, the ventricles were separated in twohalves by a short-axis section through the midportion of the heart. The upper part was immediately fixed in 4% of formaldehyde for 24 h, dehydrated with sucrose, embedded in Tissue-Tek (Sakura, United States) and frozen in liquid nitrogencooled isopentane and lower part was snap frozen in liquid nitrogen immediately after dissection. 10 μm heart sections were made using a microtome (Leica Microsystems, Nanterre, France), stained with hematoxylin and eosin for visualization of general morphology and Sirius Red for visualization of fibrosis, mounted in Eukitt (CML, France) and examined by light microscopy. Images were taken using a microscope (Leica Microsystems, Nanterre, France) equipped with a digital camera. The presence of excessive inflammation was examined by immunofluorescence using an antibody directed against vimentin (rabbit polyclonal, dilution 1:500, Progen Biotechnik, #GP53, Heidelberg, Germany). The specific binding of primary antibody was revealed using Alexa 488-conjugated secondary (dilution 1:1000, goat-anti-mouse, dilution 1:1000; Life Technologies). The slides were mounted with mowiol containing 5 μg/ml Hoescht 33,342 (Life Technologies). Images were taken using a microscope (Leica Microsystems, Nanterre, France) equipped with a digital camera.

## II.11 Statistical analyses

Analyses were conducted using GraphPad Prism 7.00 (GraphPad Software Inc., SD, United States). For one-way analysis of variance, the normality was checked using the Shapiro-Wilk normality test. If the normality of distribution assumption was not met, the nonparametric Kruskal–Wallis test was used instead of ANOVA. If a significant difference was found, then multiple comparison tests were performed to compare the different groups analyzed (Dunn's or Holm-Sidak's multiple comparison test following a nonparametric or parametric test, respectively). A p-value $\leq 0.05$ was considered significant. Values are given as the means ± standard error of the mean (SEM).

# III. Results and discussion

## III.1 Plasma phase characterization: The Ar-N$_2$ plasma showed the most energetic micro-discharges and led to the highest levels of OH radicals

### III.1.1 Electrical characterization: Mixing N$_2$ with Ar plasma increased the electrical energy of the micro-discharges

The PVDF scaffolds were treated by cold plasma in a dielectric barrier device (DBD) supplied with a carrier gas (helium or argon at 2 slm) with/without reactive gas (molecular nitrogen, 0-100 sccm). The electrical characteristics of the Ar, Ar-N$_2$, He and He-N$_2$ plasmas were investigated by measuring their voltage (V$_{pl}$) and current ($I_{pl}$). In Figure 2A, the plasma voltage remains the same whatever the gas mixture, each discontinuity corresponding to a transient gas breakdown, as highlighted in inset. These breakdowns correspond to micro-discharges that are detected as current peaks in Figure 2B. The influence of gas mixture is clearly visible on these peaks and several observations are noteworthy: 1) the amplitude of the current peaks remains in the 500-2000 mA range using argon while it falls to values between 5 mA and 50 mA using helium, 2) the overall number of current peaks is always lower with argon (typically a dozen per period) than with helium (typically more than 60 peaks per period), 3) the admixture of molecular nitrogen to argon gas does not significantly change the peaks amplitude but reduces their number, 4) its admixture to helium strongly influences the peaks amplitude although their number is decreased.

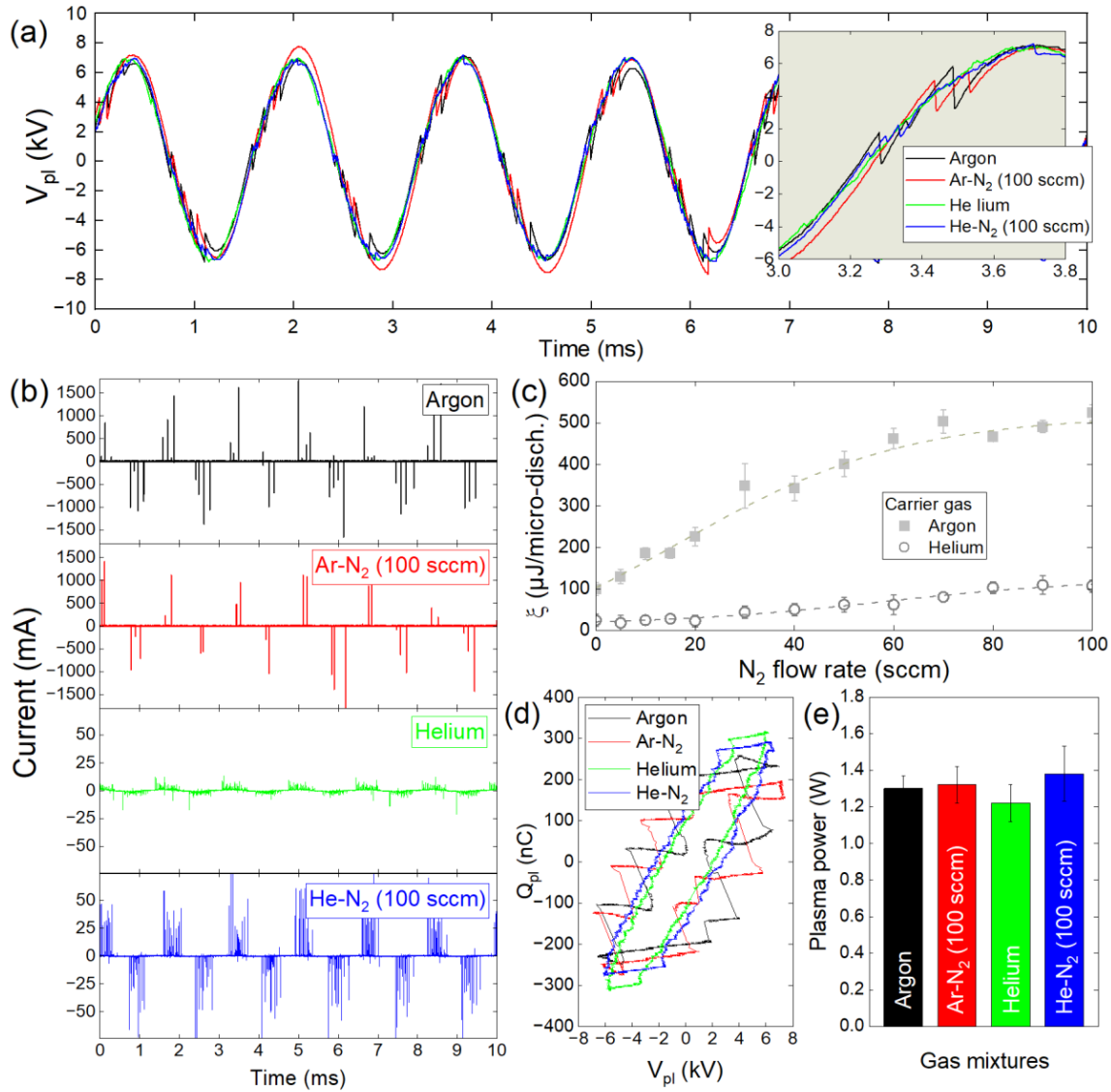

*Figure 2. Electrical characterizations of cold plasma in dielectric barrier devices supplied in Ar or He (2 slm) with/without molecular $N_2$ (0-100 sccm). (a) time profiles of plasma voltage, (b) time profiles of plasma current (c) Energy per micro-discharge versus $N_2$ flow rate, (d) Lissajous diagrams, (d) Plasma power versus gas mixtures.*

Among the four gas mixtures studied here, the Ar-$N_2$ plasma provided the highest current peaks, i.e. the micro-discharges with the highest values of electron density and therefore of electrical energy. To support this result, the average energy per microdischarge ($\xi$) is reported in Figure 2C, considering the two carrier gas with/without reactive gas. The value of $\xi$ is close to 100 $\mu J$/micro-discharge in argon while it only raises 25 $\mu J$/microdischarge in helium. Then, increasing the $N_2$ flow rate up to 100 sccm drives to $\xi$ values of only 100 $\mu J$/micro-discharge for He-$N_2$ plasma and as high as 500 $\mu J$/micro-discharge for the Ar-$N_2$ plasma. To complete this electrical characterization, the plasma power was assessed following the Lissajous method to plot the overall charge of plasma ($Q_{pl}$) as a function of plasma voltage ($V_{pl}$) in Figure 2D. The area of each closed contour multiplied by frequency (500 Hz) then gives the plasma power. As reported in Figure 2E, the plasma power is constant with a value close to 1.3W whatever the gas mixtures. All these results indicate that the nature of the gas mixture does not change the plasma power but allows to control the average energy per microdischarge, and thus to modulate the formation of active species, as demonstrated in the following paragraph.

### III.1.2 Optical characterization: The highest levels of OH radicals were obtained in Ar-N₂ plasma

The optical emission of the radiative species in He-N₂ and Ar-N₂ plasmas has been quantified using optical emission spectroscopy. The Figure 3 reports the absolute optical emission of the following species per micro-discharge as a function of the N₂ flow rate: Ar $(2p_{1/2} - 2p_{3/2})$, He $(3S_1 - 3P_1)$, OH $(A^2\Sigma^+ - X^2\Pi)$, O $(5P_{3,2,1} - 5S_2)$, N₂ $(c^3\Pi_u - b^3\Pi_g)$ and N₂⁺ $(B^2\Sigma_u^+ - X^2\Sigma_g^+)$. A first observation is that whatever the emissive species considered, the optical emission per micro-discharge is always more important with Ar-N₂ plasma (Figure 3A) than with He-N₂ plasma (Figure 3B). In the Ar-N₂ plasma, only molecular nitrogen is strongly excited: a rise in more than two decades is observed for N₂ flow rate increasing from 0 to 100 sccm. If argon metastable species are responsible for this behavior, they do not induce the production of N₂ ions because Penning ionization is not possible, as explained in section III.1.3. The OH emission slightly increases and remains at a high level (typically 10⁶ a.u.) for all admixtures of molecular nitrogen. The production of this radical results from the dissociation of water molecules which is more efficient when the N₂ flow rate is increased. Atomic oxygen species are also detected but their emission is very low in argon plasma and decreases further with the addition of molecular nitrogen.

In the He-N₂ plasma, N₂ and OH species are produced but all of their values remain almost 100 times lower than those of the Ar-N₂ plasma. Unlike argon, helium can lead to the production of high energetic metastable species to produce N₂⁺ ions and seed electrons by Penning ionization. The role of this reaction is detailed in section III.1.3.

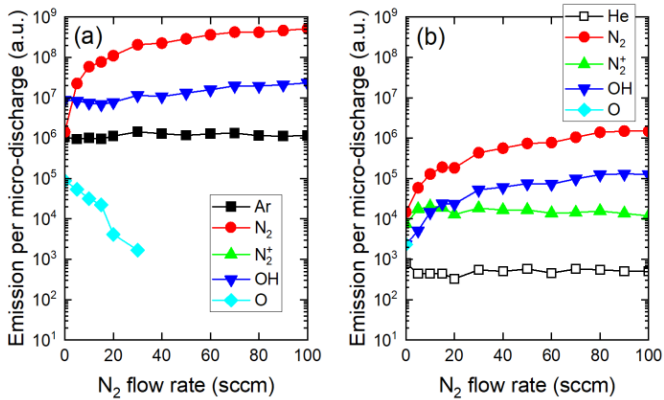

**Figure 3.** Influence of N₂ flow rate on optical emission per micro-discharge (a) Ar-N2 plasma and (b) He-N2 plasma.

### III.1.3 The carrier gas nature influenced the filamentary/ homogeneous behavior of plasma

Correlating Figure 1 and Figure 2b shows that a cold plasma is filamentary if it is generated in argon (without/with N₂) and homogeneous if it is generated in helium (without/with N₂). The formation of the emissive species detected in Figure 3 was therefore strongly related to the gas mixture composition and therefore to the plasma filamentary/homogeneous behavior which relies on two main parameters: the electron impact ionization cross sections and the seed electron density, as detailed hereafter.

In a cold plasma, the energy of the ionizing electrons depends on the value of the reduced electric field (E/ρ) whatever the nature of the carrier gas (*e.g.* helium, argon) (Chen et al., 2020). These energetic electrons collide with atoms whose characteristic radii are 31 pm for helium or 71 pm for argon. Consequently, and as supported in Figure 4a, the ionization cross sections (σ) of these species are quite different, with values of approximately $2.10^{-21}$ m² for helium and $2.10^{-20}$ m² for argon if one considers electrons with energies higher than 25 eV. Overall, the ionization cross-sections are at least 10 times lower for helium than for argon atoms and have a direct impact on electron mean free path (λe) and electron frequency (fe) (Pancheshnyi et al., 2012). λe and fe are defined by equations {1} and {2} respectively, where nn is the density of the gas atoms, ve the electron velocity and σ the ionization cross-section:

$$\lambda_e = \frac{1}{n_n . \sigma} \quad \{1\}$$
$$f_e = n_n \sigma . v_e \quad \{2\}$$

It turns out that $\lambda_e$ is almost 10 times larger in helium gas than in argon gas and that $f_e$ – which also stands for ionization frequency – is 10 times larger in argon than in helium (Yadav et al., 2008). These two reasons explain why energetic electrons can trigger primary avalanches of electrons (i.e. ionizing events) that are more numerous and operating over shorter distances in argon than in helium, hence promoting a filamentary behavior of cold plasma. The second parameter responsible for the filamentary/ homogeneous behavior is seed electron density (nseed). Cold atmospheric plasmas generated in dielectric barrier devices supplied with argon or helium commonly have a background of ambient air, *i.e.* N₂ and O₂ molecules (typically 10-100 ppm) even after a few minutes of purge (Wang et al., 2021). These so-called contaminants can interact with atoms of noble gas in a metastable state (Xᵐ) and potentially induce Penning reaction as shown in Equation {3} where molecular nitrogen is considered as contaminant. In this case, such reaction provides $N_2^{*+}$ ions as well as seed electrons, *i.e.* electrons that can be generated after gas breakdown owing to the long lifespan of the metastable species, typically a few μs.

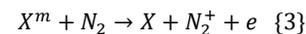

$$X^m + N_2 \rightarrow X + N_2^+ + e \quad \{3\}$$

The Penning reaction occurs only if the energy associated to Xᵐ is higher than the ionization energy of the contaminant. As reported in Figure 4b, the energy of helium metastable atoms and of argon metastable atoms is 20.0 eV and 11.6 eV respectively, while the energy required to ionize N₂ (in N₂⁺) and O₂ (in O₂⁺) is 15.5 eV and 13.6 eV respectively (Lee and Chung, 2011; Lytle et al., 2011; Jitsomboonmit et al., 2012). As a result, Penning reaction can only occur in helium gas, hence driving to important seed electron densities while in an argon plasma, nseed remains at natural level (natural cosmic radiations) and therefore negligible. For these reasons:

- In an argon plasma where $n_{seed}$ is very low, a limited number of primary avalanches is generated. Photoionization and photoemission initiate secondary avalanches which tend to converge toward the primary avalanche, driving to the development of a highly conductive filamentary channel, *i.e.* a streamer moving from an electrode to the other. This explains why argon plasma has a regime of filamentary discharge (Gherardi and Massines, 2001).
- In a helium plasma where $n_{seed}$ is elevated, many primary avalanches can be initiated. They interfere with each other, gradually forming a large positive space charge area. Consequently, the electrons resulting from the secondary avalanches do not converge anymore toward a single point. Instead, many streamers develop simultaneously; they interfere and develop as a single discharge channel covering a large area (Palmer, 1974). This explains why helium plasma has a regime of homogeneous discharge (Kogelschatz, 1992).

Now that the main physicochemical properties of cold plasma have been characterized in different gas mixtures, their effects on drop-cast PVDF scaffolds have been characterized, as well as the cyto- and biocompatibility of cardiomyocytes.

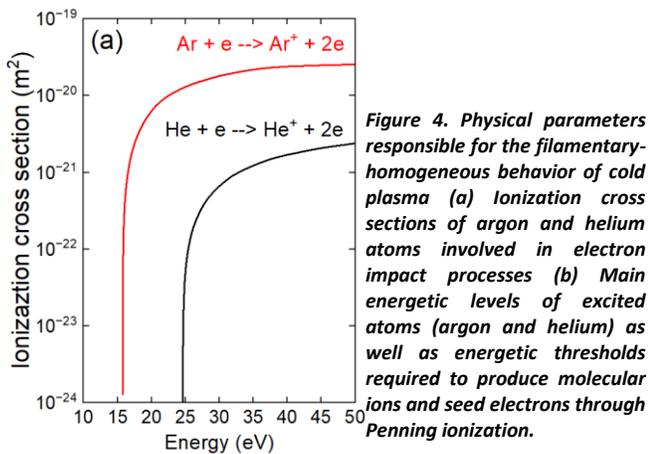

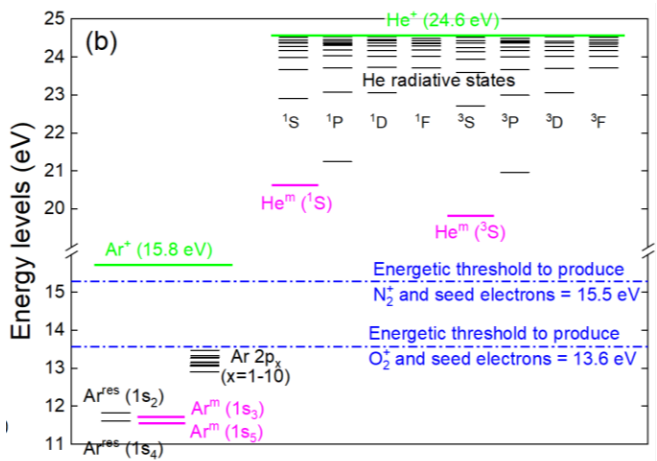

*Figure 4. Physical parameters responsible for the filamentary-homogeneous behavior of cold plasma (a) Ionization cross sections of argon and helium atoms involved in electron impact processes (b) Main energetic levels of excited atoms (argon and helium) as well as energetic thresholds required to produce molecular ions and seed electrons through Penning ionization.*

## III.2 Drop-cast polyvinylidene fluoride scaffolds exhibited globular morphology and high porosity

The drop-cast PVDF scaffolds exhibit a highly porous morphology, as shown in the SEM images in Figure 5. The globular morphology as well as the high porosity of these scaffolds, is consistent with previous results (Kitsara et al., 2019). The thicker scaffolds (5% PVDF) were used for the physicochemical and biological studies in this paper while the thinner scaffolds (3% PVDF) were not used due to their excessive brittleness and difficulties in handling. Interestingly, these scaffolds presented identical morphology on both the top and bottom sides, as shown in Figures 5 C,D respectively. This means that the drop-casting process did not affect this surface property, which is a major advantage for subsequent graftings achieved on pre-clinical models since no special care is required for marking the sides (see also Supplementary Figure S1).

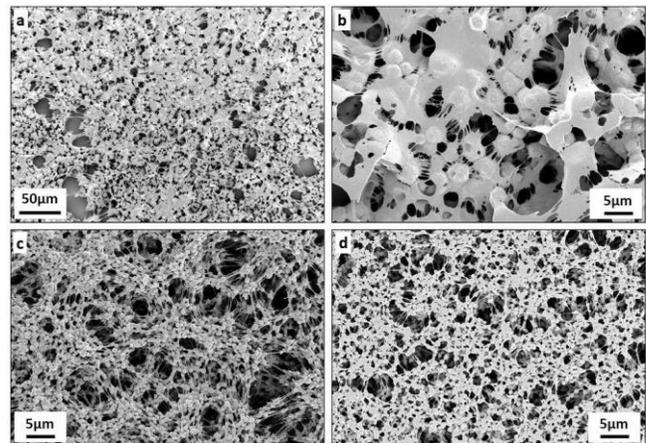

*Figure 5. Scanning electron microscope (SEM) images of drop-cast scaffolds: (a), (b) Thin PVDF obtained by 3% w/v solution. (c), (d) Thick PVDF obtained by 5% w/v solution. The images (a), (b), (c) correspond to the top side of the scaffold, while (d) shows the bottom side that is in contact with the lamella.*

## III.3 Plasma treatments hydrophilized the surface of polyvinylidene fluoride scaffolds by increasing the amount of carbon-oxygen based groups

Regardless of the plasma treatment (He or Ar with/without N$_2$ gas mixture), the PVDF surface became hydrophilic, as shown in Figure 6 which reports the shape profiles of the deposited drops as well as the corresponding water contact angles (WCAs) values. While the native surface of PVDF scaffold has a WCA value as high as 132.2°, it decreases to 52.4° and even 17.3° after Ar-N$_2$ plasma and He-N$_2$ plasma respectively. These WCA values can be correlated with the production of OH radicals within the plasma phase, as evidenced by optical emission spectroscopy in Figure 3. Unsurprisingly, the most hydrophilic states correspond to the highest emissions of OH radicals.

| Condition | Drop profile | WCA (°) |
|---|---|---|
| Control | | 132.2° (±2.1°) |
| Ar | | 115.8° (±2.4°) |
| Ar-N₂ | | 52.4° (±1.3°) |
| He | | 44.1° (±2.3°) |
| He-N₂ | | 17.3° (±2.9°) |

*Figure 6. WCA measurements on plasma-treated PVDF scaffolds.*

This gain in hydrophilicity can also be correlated with XPS analyses performed on the carbon C 1s and fluorine F 1s regions. As shown in Figure 7A, the C 1s peak of native PVDF scaffolds is only composed of carbon groups, namely -CH = CF₂-, -CF₂-CH₂-, -CF₂-CH₂-, C-C and C-H. On the contrary, the Ar-N₂ plasma treatment leads to the formation of additional carbon-oxygen based groups, as evidenced in Figure 7B and which correspond to -CFO-CH₂- and -C-C=O(OH)- at 288.5 eV, as well as -CF₂-CHO- and -CF₂-COOH at 287.1 eV. The resulting chemical surface composition of the PVDF scaffolds is reported in Table 1. Typically, while carbon concentration is 61.7% for non-treated scaffolds, this value decreases to 52.5% for Ar plasma and 51.2% for Ar-N₂ plasma. Likewise, oxygen concentration is increased from roughly 2% (native scaffolds) to approximately 9% after any plasma exposure. This is also verified on the oxygen O 1s signal with increased intensities and contributions shifts towards higher binding energies that correlate with the hydrophilicity changes (WCA analysis). Carbon bonds related to C 1s, oxygen bonds related to O 1s and CF₂ bonds related to F 1s have concentrations whose values are reported in Table 2, considering PVDF scaffolds before/after plasma treatments. While the plasma treatments decrease the concentrations of the carbon-fluorine groups (without containing oxygen atoms), the concentrations of the carbon-oxygen based groups increase. As an example, the C 1s peak indicates that the concentration of -CF₂-CH₂- decreases from 45.9% (native PVDF scaffolds) to 23.5% (PVDF scaffolds exposed to Ar-N₂ plasma). Likewise, the O 1s peak indicates that the -C=O(OH) group shows concentrations increasing from 0.8% (native PVDF scaffolds) to 4.6% (PVDF scaffolds exposed to Ar-N₂ plasma).

| Condition | $\Phi_{N2}$ (sccm) | Surf. atomic comp. | | | At. Ratio | |
|---|---|---|---|---|---|---|
| | | C 1s | F 1s | O 1s | F/C | O/C |
| No treatment (Native PVDF) | - | 61.7% | 35.7% | 2.1% | 0.58 | 0.03 |
| After Ar(N₂) plasma exposure | 0 | 52.5% | 37.4% | 10.1% | 0.71 | 0.19 |
| | 20 | 51.2% | 40.0% | 8.8% | 0.78 | 0.17 |
| After He(N₂) plasma exposure | 0 | 51.3% | 40.9% | 7.8% | 0.80 | 0.15 |
| | 20 | 51.1% | 40.1% | 8.8% | 0.78 | 0.17 |

*Table 1. Chemical surface composition of PVDF scaffolds without/with plasma treatment.*

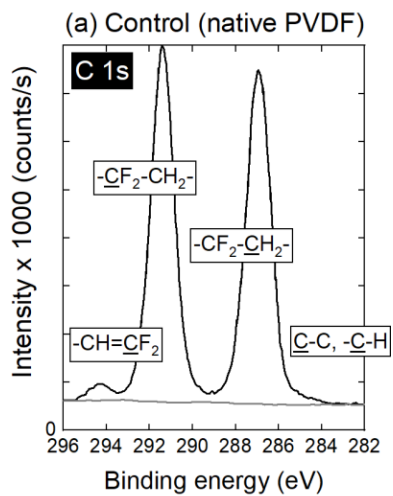

### (a) Control (native PVDF)

### (b) Ar-N₂ plasma treatment

*Figure 7. C 1s high resolution XPS spectra of (a) untreated PVDF and (b) PVDF treated with Ar-N₂ plasma ($\Phi_{N2}$ = 20 sccm).*

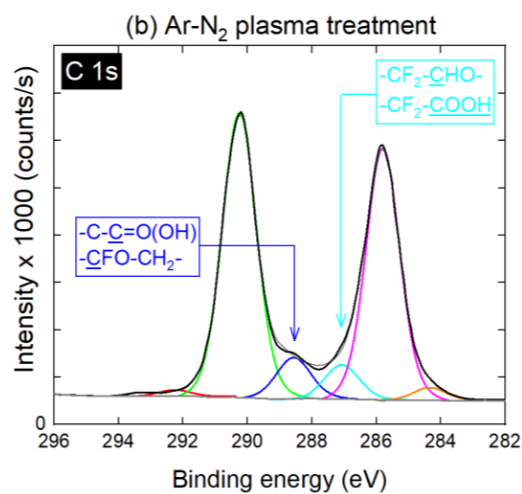

| Condition | $\Phi_{N2}$ (sccm) | C 1s | | | | | |
|---|---|---|---|---|---|---|---|
| | | -CH=CF₂ | -CF₂-CH₂ | -C-C=O(OH) -CFO-CH₂- | -CF₂-CHO- -CF₂-COOH | -CF₂-CH₂ | -C-C, -C-H |
| Native PVDF | - | (292.1 eV) 3.5% | (290.2 eV) 45.9% | (288.5 eV) 0.5% | (287.1 eV) 0.3% | (285.6 eV) 42.6% | (284.4 eV) 1.1% |
| PVDF after Ar(N₂) plasma exposure | 0 | (292.1 eV) 0.8% | (290.2 eV) 21.2% | (288.4 eV) 4.2% | (287.2 eV) 4.2% | (285.5 eV) 21.7% | (284.4 eV) 0.8% |
| | 20 | (291.9 eV) 0.4% | (290.2 eV) 23.5% | (288.5 eV) 3.1% | (287.0 eV) 2.8% | (285.7 eV) 20.4% | (284.4 eV) 1.4% |
| PVDF after He(N₂) plasma exposure | 0 | (292.1 eV) 0.6% | (290.3 eV) 23.2% | (288.6 eV) 2.4% | (287.3 eV) 2.5% | (285.9 eV) 21.5% | (284.5 eV) 1.3% |
| | 20 | (292.3 eV) 0.5% | (290.3 eV) 23.1% | (288.6 eV) 3.3% | (287.1 eV) 2.8% | (285.8 eV) 20.5% | (284.3 eV) 1.1% |

| Condition | $\Phi_{N2}$ (sccm) | O 1s | | F 1s |
|---|---|---|---|---|
| | | -C=O(OH) -CF₂-CHO- -CFO-CH₂- | -C=O(OH) | -CF₂- |
| Native PVDF | - | (532.5 eV) 0.6% | (533.8 eV) 0.8% | (687.4 eV) 4.7% |
| PVDF after Ar(N₂) plasma exposure | 0 | (532.5 eV) 5.4% | (533.8 eV) 4.8% | (687.4 eV) 36.9% |
| | 20 | (532.4 eV) 4.3% | (533.7 eV) 4.6% | (687.4 eV) 39.6% |
| PVDF after He(N₂) plasma exposure | 0 | (532.6 eV) 3.9% | (533.9 eV) 3.9% | (687.5 eV) 40.7% |
| | 20 | (532.4 eV) 5.0% | (533.8 eV) 4.1% | (687.3 eV) 39.6% |

*Table 2. Concentrations of the carbon bonds related to C 1s, the oxygen bonds related to O 1s and the CF₂ bond related to F 1s for PVDF scaffolds before/after plasma treatments. Concentrations are expressed in %. Values in parenthesis correspond to the binding energies expressed in eV.*

## III.4 Plasma-treated PVDF scaffolds improved cytocompatibility, adhesion and morphology of cardiomyocytes

Well-known and well-characterized primary cardiomyocytes from newborn rats were isolated following the protocol described in Section II.5. Six days after their cellularization onto non-treated and plasma-treated PVDF scaffolds, these primary cardiomyocytes were immune-stained with a sarcomeric marker (Tandon et al., 2013) to evaluate their adhesion, cytocompatibility and morphological changes using fluorescence microscopy.

*Figure 8. Cellularization of the plasma-treated PVDF scaffolds with newborn rat primary cardiomyocytes. Results are obtained six days after seeding cardiomyocytes on (a and a') non-treated PVDF, (b and b') He plasma-treated PVDF, (c and c') He-N2 plasma-treated PVDF, (d and d') Ar plasma-treated PVDF, (e and e') Ar-N2 plasma-treated PVDF. The cardiomyocytes appear green (due to their immunolabeling by antibody against α-actinin) while the nuclei appear in blue (due to Hoescht 33342 staining). White dashed framings show a higher magnification area. (f) Cardiomyocyte surface area and perimeter, cardiomyocyte length-to-width ratio and roundness index were measured using ImageJ software. Values are given as means ± standard error of the mean. # indicate a significant difference compared to the non-treated PVDF scaffold. \* indicate a significant difference compared to the He-N2 PVDF scaffold. Statistical analyses were realized using non-parametric Kruskal-Wallis test. If a significant difference was found, Dunn's post hoc tests for multiple comparisons were used.*

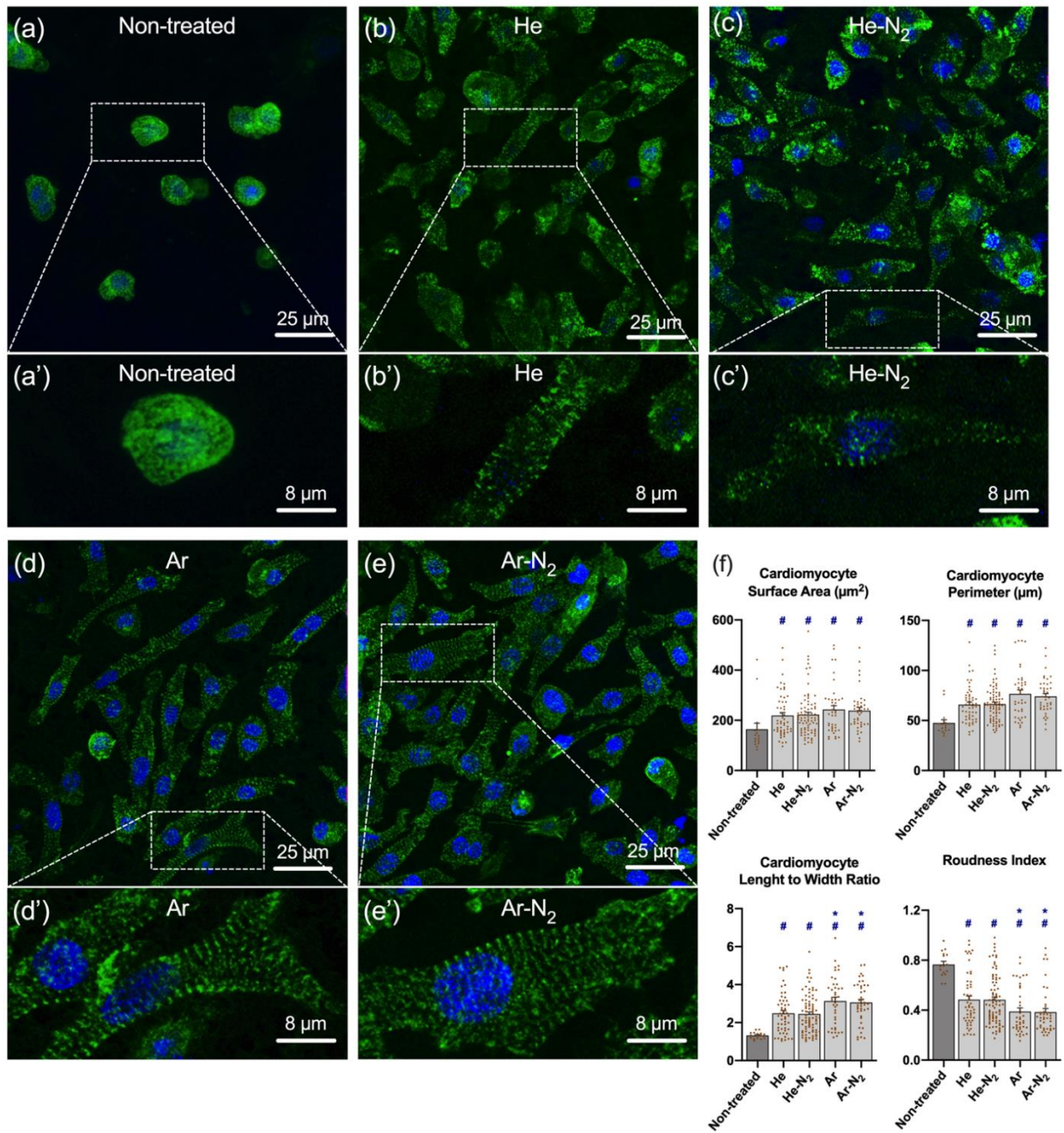

As shown in Figure 8, a significant increase in the number of cardiomyocytes was observed on all plasma-treated PVDF scaffolds compared to the non-treated ones. Hence, the Figure 8A reveals that only few cardiomyocytes adhered on the non-treated PVDF scaffolds, these cells being stunted/retracted on themselves and isolated from each other. On the contrary, the cardiomyocytes upon culture on all the plasma-treated PVDF scaffolds show a well-organized sarcomeric structure, as observed by α-actinin immunostaining from Figures 8B–E. It is expected that cardiomyocytes seeded on a scaffold will immediately adhere to its surface if the latter presents adequate physico-chemical properties (i.e., chemical functionalization, morphology). Then, during the first days of culture, a considerable increase in their sizes is expected, as well as a change in their morphology (from circular to rectangular), resulting in an increase in the length to width ratio, with a typical value of 7:1 for adult cardiomyocytes (Yang et al., 2014; Ribeiro et al., 2015).

For these reasons, we determined the ability of the newborn rat primary cardiomyocytes to grow on non-treated and plasma-treated PVDF scaffolds by measuring the four following parameters after 6 days of culture: cardiomyocyte surface area, cardiomyocyte perimeter, length-to-width ratio and roundness index. It should be noted that similar morphological analyses to explore the behavior of cardiomyocytes on different surfaces have been widely documented in the literature (Yang et al., 2014; Li et al., 2021). As represented in Figure 8F, we noticed an increase in the cardiomyocyte size (surface and perimeter), the cardiomyocyte length-to-width ratio of the cells on all plasmatreated PVDF in comparison to non-treated PVDF scaffolds after 6 days of culture. In contrast, we noticed an important decrease in the roundness index with a value as high as $0.77 \pm 0.02$ for non-treated PVDF scaffolds and as low as $0.39 \pm 0.03$ for Ar, $0.38 \pm 0.03$ for Ar-$N_2$, $0.48 \pm 0.03$ for He and $0.48 \pm 0.02$ for He-$N_2$ plasma-treated PVDF scaffolds.

These findings suggest that cardiomyocytes can easily adhere and spread, as well as exhibit a physiological shape on the plasma-treated PVDF scaffolds. As shown in Figure 8F, the cardiomyocyte length-to-width ratio and roundness index in cardiomyocyte cultured on Ar and Ar-$N_2$-treated PVDF were significantly different compared to the other plasma treatments. Cardiomyocyte seeded on Ar and Ar-$N_2$-treated PVDF exhibit a less circular and more rectangular morphology than cardiomyocytes seeded on He and He-$N_2$ treated PVDF, indicating that Ar- and Ar-$N_2$-treated PVDF are more suitable for cardiomyocyte culture. It would be interesting to validate these results using human cardiomyocytes derived for example from pluripotent stem cells to demonstrate the interest of this scaffold as well as the importance of plasma-treatment for cardiomyocyte culture.

## III.5 Plasma-treated PVDF scaffolds increased the expression of genes related to adhesion and cardiac function of cardiomyocytes

The efficacy of plasma treatment was also supported by real time PCR analysis which demonstrated an increased expression of the genes related to adhesion and cardiac function of cardiomyocytes.

For further analysis, adhesion index and cardiac index were constructed, as described in Section II.8. As shown in Figure 9A, whatever the plasma treatment, no significant difference was observed amongst the samples in terms of cell adhesion, although they all exhibited higher adhesion in comparison to the non-treated PVDF. It turns out that cell adhesion is not directly correlated with plasma induced surface hydrophilization; this observation is consistent with a recent work pointing out that surface wettability measured by drop shape analysis is not always a relevant predictor of cell adhesion to scaffolds (Alexander and Williams, 2017). For the genes related to the cardiomyocyte function, the cardiac index in cardiomyocytes cultured on Ar-$N_2$ treated PVDF was significantly higher compared to the other treatments, as clearly evidenced in Figure 9B. This result can be explained by considering that the drop-cast PVDF scaffolds have a high porosity, allowing the radicals generated by plasma not to remain on the apparent surface of these scaffolds but to diffuse in-depth and through the interstices. Among the four gas mixtures investigated to generate plasma, the microdischarges from the Ar-$N_2$ plasma were the most energetic (up to 500 µJ/micro-discharge according Figure 2C) and produced the highest densities of OH radicals (Figure 3).

These two physico-chemical parameters seem responsible for the enhanced grafting of carbon-oxygen based groups on the surface of the PVDF scaffolds. This explanation is also supported by many works where the diffusion of oxygenated species was evidenced through different polymer films like Nafion or LDPE (Li et al., 2011; Abou Rich et al., 2015).

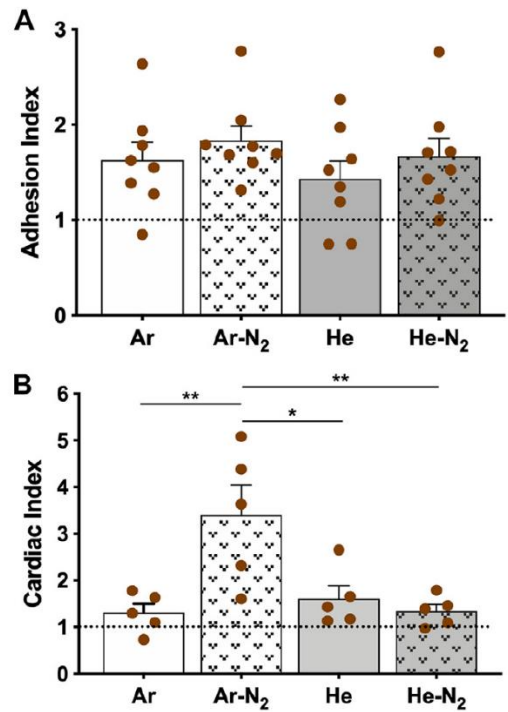

*Figure 9. Adhesion and cardiac index. (A) Adhesion index, (B) Cardiac index. The scorewas calculated using three independent experiments. Non-treated surface values are set at 1.0 (dotted line). Values are given as means ± standard error of the mean. Statistical analyses were realized using parametric ANOVA test. If a significant difference was found Holm-Sidak's post hoc test for multiple comparisons were used). \*p < 0.05, \*\*p < 0.01.*

## III.6 Plasma-treated PVDF implants were bio-compatible with living mouse hearts

The biocompatibility of the PVDF scaffolds was studied in a cardiac environment, after implantation of acellular scaffolds on the surface of the heart of healthy animals as previously described in our studies (Fiamingo et al., 2016; Joanne et al., 2016; Flaig et al., 2020; Domengé et al., 2021). We first evaluated the integrity of the nontreated PVDF scaffolds 7 and 28 days after implantation (Figure 10).

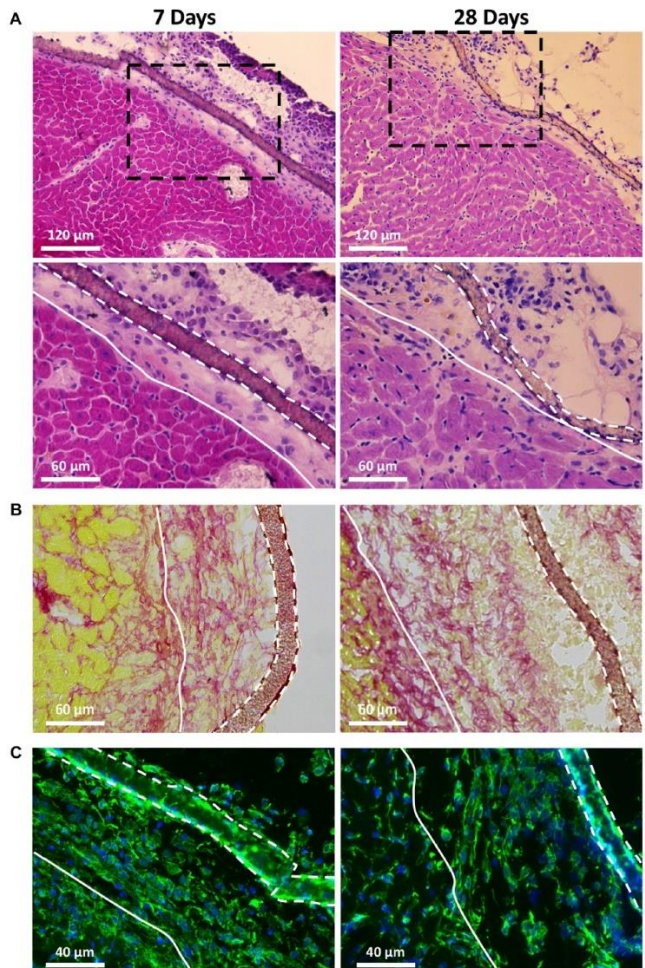

**Figure 10. Biocompatibility of the PVDF scaffolds after implantation on the surface of the heart of healthy mice. The residual scaffolds and grafted area were visualized after 7 days and 28 days of implantation of non-treated PVDF scaffold by hematoxylin-eosin staining (A); Sirius Red staining (B) and vimentin immunostaining (C). Black dashed framings in (A) show a higher magnification area (of the lower panel). White solid lines show the border between the grafted area and heart, while white dashed lines represent residual mats on the surface of the heart. Note the absence of multinucleated giant cells, exacerbated fibrosis and inflammation in the grafted area.**

At these two dates, Figure 10A shows that the non-treated PVDF scaffolds were visible and well-integrated on the surface of the ventricle without any visible sign of degradation. More specifically, this figure shows that the grafted area had no exuberant fibrosis and no multinucleated giant cells: a common feature of granulomas that could have developed in case of various inflammatory reactions. The biocompatibility of PVDF scaffold was also supported by Sirius red staining (Figure 10B) and vimentin immunostaining (Figure 10C). Sirius red staining which is a commonly used histological technique to visualize collagen, revealed no excessive fibrosis around the grafted PVDF at either 7 or 28 days post-implantation. This result was also confirmed by immunostaining for vimentin, a marker of fibroblasts and some inflammatory cells. In response to inflammatory stimuli, vimentin expression increased rapidly because it belongs to the family of "immediate early genes," a group of genes that are rapidly activated (Lilienbaum and Paulin, 1993; Li et al., 2020; Paulin et al., 2022). As shown in Figure 10C, vimentin positive cells are homogenously located between surface of the heart and the grafted area, without clustering around the scaffold. After evaluating the biocompatibility of the native PVDF scaffold, we were interested in assessing the biocompatibility of the plasma-treated PVDF scaffolds.

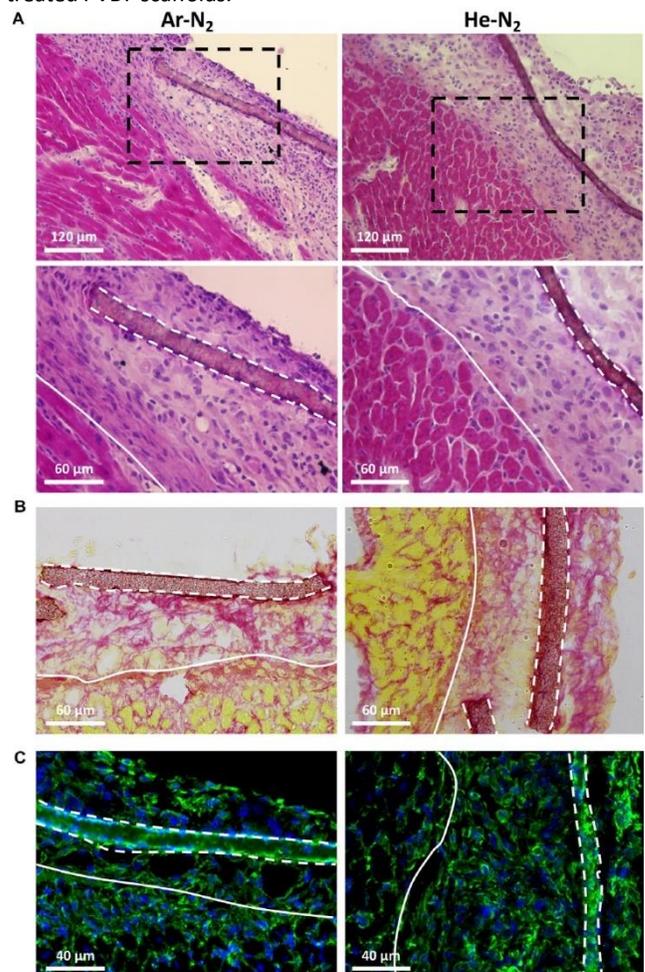

**Figure 11. Biocompatibility of the plasma treated PVDF scaffolds after implantation on the surface of the heart of healthy mice. The residual scaffolds and grafted area were visualized after 7 days of implantation of Ar-N₂ or He-N₂ treated PVDF scaffold by hematoxylin-eosin staining (A); Sirius Red staining (B) and vimentin immunostaining (C). Black dashed framings in (A) show a higher magnification area (of the lower panel). White solid lines show the border between the grafted area and heart, while white dashed lines represent residual mats on the surface of the heart. Note the absence of multinucleated giant cells, exacerbated fibrosis and inflammation in the grafted area.**

Figure 11 shows that 7 days after implantation, the plasma-treated PVDF scaffolds were successfully integrated on the heart tissue, whether after Ar-$N_2$ or He-$N_2$ plasma treatments. As demonstrated with non-treated PVDF scaffolds, no exuberant fibrosis and no multinucleated giant cells were observed in the grafted area, only mononuclear cell infiltration was observed, hence demonstrating the absence of a foreign body reaction and the biocompatibility of these scaffolds. Moreover, plasma-treated PVDF scaffolds did not induce excessive activation of vimentin positive cells around the scaffold, 7 days after implantation on the surface of the heart. This result is consistent with experimental works where the biocompatibility of PVDF scaffolds was effectively promoted to fourfold by applying $O_2$ plasma treatments while other works clearly stated that PVDF could not produce any cellular immune response (T-lymphocytes, macrophages, and neutrophils) (Klink et al., 2011; Mangindaan et al., 2012).

# IV. Conclusion

In this work, we have demonstrated that cold atmospheric plasmas can be an efficient approach to modify the surface chemistry of drop-cast PVDF scaffolds for cardiac tissue engineering. Specifically, the Ar-$N_2$ plasma generated highly energetic micro-discharges with high levels of OH radicals, leading to the formation of carbon-oxygen based groups on the surface of PVDF scaffolds. The high porosity of these scaffolds suggested that the radicals were also able to diffuse to the subsurface and create deeper functionalization. Hence, four types of carbon-oxygen based groups were identified as being responsible for the improved wettability properties of the PVDF scaffolds.

The better cytocompatibility of these scaffolds was demonstrated once cellularized with newborn rat primary cardiomyocytes. Indeed, higher adhesion were evidenced, as well as a change in cardiomyocyte morphology, characterized by a well-organized sarcomeric structure. In addition, real-time PCR showed that the expression of the genes related to adhesion and cardiac function were increased in the case of the plasma-treated PVDF scaffolds. Finally, an in vivo study was carried out on healthy murine models to verify that these scaffolds did not increase inflammation once implanted on hearts. Since no toxic response was observed and since PVDF is known to not produce immune responses, plasma-treated PVDF scaffolds appear as promising biocompatible candidates for cardiac tissue engineering. This work is an essential step before conducting further experiments on human cells, i.e. cardiomyocytes derived from induced pluripotent stem cells. It also paves the way for clinical studies to confirm the long-term biocompatibility of the drop-cast PVDF scaffolds treated by Ar-$N_2$ plasma.

# V. Ethics statement

All procedures were performed in accordance with national and European legislations, in conformity with the Public Health Service Policy on Human Care and Use of Laboratory Animals under the license B75.13.20. All animal studies were approved by our institutional Ethics Committee "Charles Darwin" (Permit number: 4370) and conducted according to the French and European laws, directives, and regulations on animal care (European Commission Directive 86/609/EEC).

# VI. Author contributions

# VII. Funding

This work was supported by the LabEx REVIVE (ANR-10-LABX-73). Also, this work has been sponsored by the Ile-de-France Region in the framework of Respore, the Île-de-France network of Excellence in Porous Solids. MK acknowledges personal funding from both LabEx REVIVE and DIM Respore too. J-SV is supported by a PhD fellowship from Fonds Marion Elizabeth Brancher. OA is supported by the AFM-Téléthon (contract numbers: 21833 and 22142).

# VIII. Acknowledgments

The authors acknowledge IMPC from Sorbonne University (Institut des Matériaux de Paris Centre, FR CNRS 2482) and the C'Nano projects of the Region Ile-de-France, for Omicron XPS

# IX. Conflict of interest

The authors declare that the research was conducted in the absence of any commercial or financial relationships that could be construed as a potential conflict of interest.

# X. Supplementary material

Table S1. Primers sequences used in this study

| Gene | Forward (5'-3') | Reverse (5'-3') |
|---|---|---|
| Actn2 | AGACATGGGCTTACGGCAAA | GCTGCAATCTGTTCCACACG |
| Myh7 | CAACCTGTCCAAGTTCCGC | TACTCTTCATTCAGGCCCTTGG |
| Pln | GCTCCCAGACTTCACACAAC | TCTCCTTTTAGGAGGCCTTGG |
| Atp2a2 | TCTGCTTGTCCATGTCCCTT | ATTGCAGGCTCCAGGTAGTT |
| Cryab | GGAGAGCACCTGTTGGAGTC | CATACGCATCTCTGAGAGCCC |
| Cdh2 | CACCCGGCTTAAGGGTGATT | CGATCCTGTCTACGTCGGTG |
| Cdh13 | TGGTCAAGCCCCTGGACTAT | ACCATCATGGGGTCTGGGTA |
| Col1a1 | TGTGCCTCAGAAGAACTGGT | CGCTTCCATACTCGAACTGG |
| ECM1 | ATAAAGACCCACCCCCACTC | TCCACAGAGATGGTCCATGA |
| ELN | GCTAAATACGGAGCAGCAGG | TACTCCACCAGGAACACCAC |
| Lama4 | TCACCACACCGATGGCTAAC | TGAGGTTTCTCACTGCGTCC |
| Lamb2 | GTCTTCGCTGTGACCACTGT | AACCAGCAATGCACCTCTCA |
| Lamc1 | TTCTACAACCTGCAGAGCGG | TGGTGACAGTCGCAAGGTTT |
| Actb | AGATCAAGATCATTGCTCCTCCT | AAGGGTGTAAAACGCAGCTC |
| B2m | TGAATTCACACCCACCGAGA | TACATGTCTCGGTCCCAGGT |
| Nubp1 | CCCAAGTGCAAGAGAGAGTC | ACTTTGCCCAGAAGAGGGAT |

Table S2. Fold changes of mRNA expression of the genes used to construct the cardiac and the adhesion index.

Non-treated surface values are set at 1.0. Values are given as the means ± standard error of the mean (SEM).

| Gene | Index | Ar | Ar-N$_2$ | He | He-N$_2$ |
|------|-------|-----|----------|-----|----------|
| *Actn2* | Cardiac index | 1.30 ± 0.53 | 3.63 ± 1.90 | 1.65 ± 1.31 | 0.98 ± 0.50 |
| *Atp2a2* | | 1.10 ± 0.40 | 5.08 ± 2.59 | 1.43 ± 1.10 | 1.09 ± 0.53 |
| *Cryab* | | 0.73 ± 0.27 | 1.61 ± 0.85 | 1.14 ± 0.44 | 1.46 ± 0.52 |
| *Myh7* | | 1.78 ± 0.55 | 4.38 ± 2.33 | 2.66 ± 1.28 | 1.79 ± 0.65 |
| *Pln* | | 1.64 ± 0.49 | 2.32 ± 1.49 | 1.18 ± 0.95 | 1.39 ± 0.87 |
| *Cdh13* | Adhesion index | 0.85 ± 0.47 | 1.78 ± 0.36 | 0.75 ± 0.40 | 1.00 ± 0.35 |
| *Cdh2* | | 2.65 ± 0.82 | 2.78 ± 0.81 | 1.98 ± 0.84 | 2.78 ± 0.61 |
| *Col1a1* | | 1.78 ± 0.71 | 1.79 ± 1.10 | 1.35 ± 0.64 | 1.72 ± 0.53 |
| *Ecm1* | | 1.63 ± 0.44 | 1.61 ± 0.76 | 1.65 ± 0.94 | 1.43 ± 0.34 |
| *Eln* | | 1.56 ± 0.47 | 1.32 ± 1.06 | 0.75 ± 0.37 | 1.23 ± 0.63 |
| *Lama4* | | 1.94 ± 0.66 | 1.70 ± 0.19 | 2.27 ± 0.95 | 1.53 ± 0.11 |
| *Lamb2* | | 1.27 ± 0.34 | 2.05 ± 1.27 | 1.19 ± 0.73 | 1.71 ± 1.00 |
| *Lamc1* | | 1.39 ± 0.15 | 1.68 ± 0.57 | 1.52 ± 0.42 | 1.98 ± 0.26 |

Table S3. Fold changes of mRNA expression of the genes used to construct the cardiac and the adhesion index before normalization.

Values are given as the means ± standard error of the mean (SEM).

| Gene | Index | Non-treated | Ar | Ar-N$_2$ | He | He-N$_2$ |
|------|-------|-------------|-----|----------|-----|----------|
| *Actn2* | Cardiac index | 4.33 ± 0.53 | 5.63 ± 2.28 | 15.72 ± 8.24 | 7.16 ± 5.68 | 4.23 ± 2.16 |
| *Atp2a2* | | 10.03 ± 1.54 | 11.05 ± 3.97 | 50.99 ± 26.01 | 14.37 ± 11.09 | 10.99 ± 5.36 |
| *Cryab* | | 5.98 ± 0.85 | 4.37 ± 1.64 | 9.61 ± 5.09 | 6.79 ± 2.64 | 8.76 ± 3.10 |
| *Myh7* | | 1.22 ± 0.15 | 2.18 ± 0.67 | 5.35 ± 2.84 | 3.24 ± 1.56 | 2.19 ± 0.80 |
| *Pln* | | 1.18 ± 1.03 | 1.93 ± 0.58 | 2.73 ± 1.76 | 1.39 ± 1.12 | 1.64 ± 1.03 |
| *Cdh13* | Adhesion index | 0.57 ± 0.01 | 0.48 ± 0.27 | 1.01 ± 0.20 | 0.43 ± 0.23 | 0.57 ± 0.20 |
| *Cdh2* | | 0.33 ± 0.10 | 0.87 ± 0.27 | 0.92 ± 0.27 | 0.65 ± 0.28 | 0.92 ± 0.20 |
| *Col1a1* | | 13.71 ± 1.48 | 24.44 ± 9.72 | 24.51 ± 15.01 | 18.47 ± 8.74 | 23.54 ± 7.28 |
| *Ecm1* | | 0.03 ± 0.01 | 0.05 ± 0.01 | 0.05 ± 0.03 | 0.06 ± 0.03 | 0.05 ± 0.01 |
| *Eln* | | 1.24 ± 0.25 | 1.92 ± 0.58 | 1.63 ± 1.31 | 0.93 ± 0.46 | 1.51 ± 0.78 |
| *Lama4* | | 0.08 ± 0.02 | 0.16 ± 0.06 | 0.14 ± 0.02 | 0.19 ± 0.08 | 0.13 ± 0.01 |
| *Lamb2* | | 0.28 ± 0.07 | 0.36 ± 0.10 | 0.58 ± 0.36 | 0.34 ± 0.21 | 0.48 ± 0.28 |
| *Lamc1* | | 0.67 ± 0.11 | 0.93 ± 0.10 | 1.13 ± 0.39 | 1.02 ± 0.28 | 1.33 ± 0.17 |

Supplementary Figure S1. Scanning electron microscope (SEM) images of drop-cast scaffolds

(a) Native PVDF scaffold obtained by 5% w/v solution, (b) PVDF scaffold after Ar plasma exposure, (c) PVDF scaffold after He plasma exposure, (d) PVDF scaffold after Ar-N$_2$ plasma exposure, (e) PVDF scaffold after He-N$_2$ plasma exposure. The images (b), (c), (d) and (e) correspond to the side exposed to plasma.

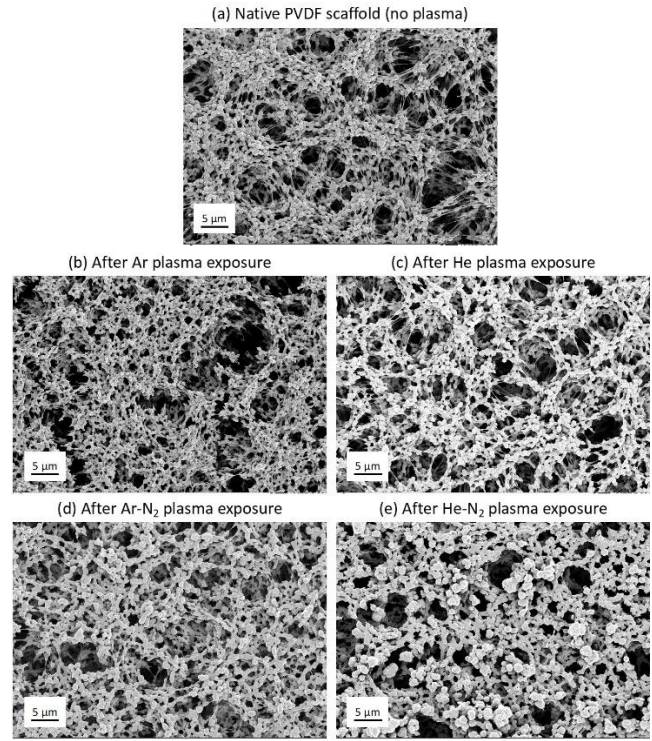

(a) Native PVDF scaffold (no plasma)

(b) After Ar plasma exposure

(c) After He plasma exposure

(d) After Ar-N$_2$ plasma exposure

(e) After He-N$_2$ plasma exposure